# TractGraphCNN: Anatomically Informed Graph CNN for Classification Using Diffusion MRI Tractography

*Yuqian Chen[1,2], Fan Zhang[1], Leo R. Zekelman[1], Tengfei Xue[1,2], Chaoyi Zhang[2], Yang Song[3], Nikos Makris[1], Yogesh Rathi[1], Weidong Cai[2], Lauren J. O'Donnell[1]*

[1] Harvard Medical School, MA, USA
[2] The University of Sydney, NSW, Australia
[3] The University of New South Wales, NSW, Australia

## ABSTRACT

The structure and variability of the brain's connections can be investigated via prediction of non-imaging phenotypes using neural networks. However, known neuroanatomical relationships between input features are generally ignored in network design. We propose TractGraphCNN, a novel, anatomically informed graph CNN framework for machine learning tasks using diffusion MRI tractography. An EdgeConv module aggregates features from anatomically similar white matter connections indicated by graph edges, and an attention module enables interpretation of predictive white matter tracts. Results in a sex prediction testbed task demonstrate strong performance of TractGraphCNN in two large datasets (HCP and ABCD). Graphs informed by white matter geometry demonstrate higher performance than graphs informed by gray matter connectivity. Overall, the bilateral cingulum and left middle longitudinal fasciculus are consistently highly predictive of sex. This work shows the potential of incorporating anatomical information, especially known anatomical similarities between input features, to guide convolutions in neural networks.

## 1. INTRODUCTION

The human brain's white matter (WM) fiber tract connections have important inter-individual variability, with implications for understanding neurodevelopment and disease [1]. Recently, brain variability is studied by predicting non-imaging phenotypes from high-dimensional neuroimaging data using machine learning [2]. Many aspects of such machine learning methods are active areas of research (e.g. multiple modalities [3], comparison of methodology [4], and interpretation [2], [4]). However, we find relatively fewer studies of tailored network design that can leverage neuroanatomical knowledge. Here we investigate deep neural networks informed by the anatomy and geometry of the brain's WM structure.

A few studies have aimed to develop dedicated neural networks for analyses of the brain's structural connections. The BrainNETCNN [5] includes novel convolutional filters that improve performance [6] by handling the topology of connectivity matrices (where each row or column corresponds to a gray matter (GM) region or node, and entries or edges in the matrix indicate connectivity strengths between GM regions). Other approaches apply graph convolutional neural networks to connectivity matrices, e.g. [7]. However, the above classes of methods are restricted to the anatomical information contained in the row and column structure of the connectivity matrix, and they cannot leverage any additional anatomical information to inform network convolutions.

We hypothesize that the performance of deep learning can be enhanced by incorporating information about *anatomical neighborhoods* of WM connections with similar geometry and connectivity. To encode neighborhood relationships, we adopt the popular EdgeConv neural network module originally designed for the Dynamic Graph CNN (DGCNN) [8], and we use it to construct static graphs informed by brain anatomy. Following two major approaches to study the brain's structural connectivity [9], we investigate 1) *white-matter-centric* graphs (WMG) that define neighborhoods according to fiber tract geometry, and 2) *gray-matter-centric* graphs (GMG) that define neighborhoods according to connected gray matter regions.

To focus our project, we choose a testbed problem of sex prediction. While this problem is not straightforward [10], [11], sex is known to be an important source of WM variability [12]. Many studies have investigated sex prediction [6], [13]–[16] using microstructure and/or connectivity features from quantitative diffusion MRI (dMRI) tractography [9]. Both microstructure (e.g. fractional anisotropy, FA [6]) and connectivity (e.g. number of streamlines, NoS [14]) features provide good prediction performance. However, NoS is affected by intracranial volume, a common confound in sex prediction [10]. In this work we therefore utilize FA and a percentage of streamlines that is normalized to reduce the effect of brain size. We quantify these features using an anatomically curated, atlas-based WM fiber cluster parcellation that is consistent across datasets, acquisitions, and the human lifespan [17]. Importantly for this study, a description of the WM geometry and GM connectivity of each fiber cluster is provided in the ORG atlas [17].

In this study, we propose an anatomically-informed graph CNN framework, called TractGraphCNN, to leverage neuroanatomical knowledge for sex prediction based on cluster-wise WM features from dMRI tractography. The main contributions of this study are as follows. First, for the first time, we model the anatomical relationship between clusters as a graph, informed by WM geometry and GM connectivity information. Second, we integrate EdgeConv modules into our framework to extract features from anatomically similar clusters to improve performance of sex prediction. Finally, our framework is able to identify important WM tracts for sex classification by leveraging an attention module. We evaluate our method on two large-scale datasets of children and healthy young adults.

## 2. METHODS

Fig. 1 gives an overview of our proposed TractGraphCNN method. First, WM features are extracted from dMRI tractography data (Sec. 2.1), resulting in two features for each cluster. Second, we build a graph to model the relationship between clusters, indicated by WM geometry or GM connectivity information (Sec. 2.2). Third, the built graph is input to the proposed TractGraphCNN framework (Sec. 2.3) for sex classification. The framework aggregates information from connecting clusters in the graph via EdgeConv modules. An attention module is adopted to enable interpretation of important tracts that are predictive for sex classification.

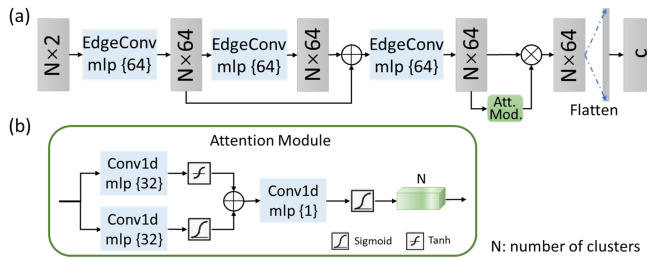

Fig. 1. (a) Overall pipeline of TractGraphCNN. (b) Network structure of the attention module.

### 2.1 dMRI datasets and feature extraction

*2.1.1 Adolescent Brain Cognitive Development (ABCD)*
This study utilized dMRI data of 9342 young children (age 9-11) from the large-scale, multi-site ABCD dataset [18]. We harmonized the dMRI data across 21 acquisition sites to remove scanner-specific biases while preserving inter-subject biological variability [19], [20]. Of all subjects, 4879 (52.2%) are males and 4463 (47.8%) are females. 7473 subjects (80%) are used for training the neural network, while 1869 (20%) are used for testing.

*2.1.2 Human Connectome Project (HCP)*
We also conducted experiments on a dataset of 964 subjects (age 22-37) from the Human Connectome Project, a large multimodal dataset composed of healthy young adults [21]. Of all subjects, 443 are male (45.6%) and 521 are female (54.4%). 772 subjects (80%) are used for training and 192 (20%) subjects are used for testing.

*2.1.3 White matter fiber cluster features*
Two-tensor unscented Kalman filter tractography (UKFt) [22] via SlicerDMRI [23], [24] was applied to obtain whole brain tractography from the dMRI data. Tractography was parcellated with an anatomically curated cluster atlas and a machine learning approach that has been shown to consistently identify WM tracts across the human lifespan [17]. For each subject, 953 expert-curated clusters categorized into 75 WM tracts were obtained. Importantly, cluster IDs are assigned according to the atlas and correspond across subjects (e.g. cluster #1 corresponds across all subjects and datasets studied). Statistical microstructure measurements were then computed for each cluster. We adopted two measurements for the task: fractional anisotropy (FA) and percentage of streamlines (PoS). FA of the cluster is computed as the mean FA across all streamline points within the cluster. The PoS of a cluster is calculated as the number of streamlines of the cluster divided by the total number of streamlines across all clusters of the subject, to reduce the effect of brain size. For each subject, this resulted in an input feature matrix of size 2x953. For absent clusters due to individual anatomical variation, we set features to zero. Finally, a min-max normalization was performed on the input feature matrix for FA and PoS individually.

### 2.2 Anatomically informed graph construction

We propose to build graphs such that edges connect neighboring fiber clusters with similar anatomy. Each cluster is represented as a node in the graph with cluster-wise WM features as node features.

*2.2.1 Fiber tract geometry informed graph*
The first type of graph (WMG) proposed in our study is based on WM tractography fiber geometric similarity, a well-established concept in the field of fiber clustering [9]. Specifically, we first compute the geometric distance between each pair of fiber clusters in the ORG atlas, which is measured as the mean of the pairwise fiber distances (the popular mean closest point fiber distance is used [25]) between the two fiber clusters. A low distance between two clusters represents a high similarity in terms of WM anatomy. Then, for each cluster, we choose the top $k$ ($k$ =20 is used in our study following the default setting in DGCNN) clusters with the lowest geometric distances as neighbors, and edges are placed in between for graph construction.

*2.2.2 Cortical and subcortical connectivity informed graph*

The second type of graph (GMG) proposed in our study is based on GM regions to which the fiber clusters connect. Specifically, for each cluster, we first identify its connected Freesurfer GM regions. The ORG atlas provides the percentage of streamlines from each cluster that intersect each Freesurfer region [26]. We leverage this information to identify the top two FreeSurfer regions most commonly intersected by the streamlines of each cluster. The neighborhood of a cluster is then defined as the set of clusters with at least one top Freesurfer region in common, and edges are placed in between for graph construction.

## 2.3 Network architecture

The overall architecture of our TractGraphCNN framework is shown in Fig. 1. TractGraphCNN extends the 1D CNN model [15] for group classification using fiber cluster features with two innovative improvements. First, we replace the 1D convolutional layers in the original model with EdgeConv layers [8] to utilize the information of anatomically neighboring clusters (Sec. 2.3.1). Second, we add a gated attention module [27] in the network that can assess the importance of each cluster to enable result interpretation (Sec. 2.3.2).

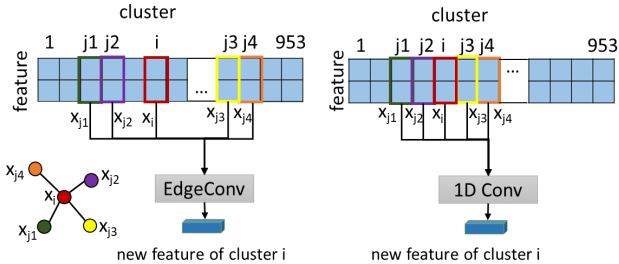

Fig. 2. Graphic illustration of the usage of EdgeConv to leverage fiber cluster neighborhood information, with comparison to the standard 1D convolutional layer.

*2.3.1 1D CNN with EdgeConv*

Fig. 2 illustrates the use of EdgeConv in TractGraphCNN to aggregate information from neighboring graph nodes representing fiber clusters. EdgeConv was proposed in the popular DGCNN method to capture the local geometric structure of point clouds [8]. The basic idea of EdgeConv is to use a learnable fully-connected layer to compute an edge feature of two neighboring nodes $x_i$ and $x_j$ based on their input features. Then, the output of EdgeConv is calculated by aggregating the edge features with max-pooling. This learning process enables dynamic update of graph structure by recomputing distances of points in the feature space. In our application, because the anatomical relationships between fiber clusters do not change, we maintain a static graph structure across layers by using the same graph structure across all EdgeConv layers without recomputing distances between node features. In addition, after feature extraction, we do not use the max pooling operation as in traditional Graph CNNs, but instead we retain the flatten operation in the 1D CNN model [15] to preserve the information about cluster correspondence across subjects.

*2.3.2 Interpretation of important tracts*

For the purpose of interpretation, it is important to identify important WM connections for the task of sex classification. To achieve this, we improve our neural network by adding an attention mechanism using the popular gated attention module from [27]. The attention module (Fig. 1(b)) is composed of two parallel fully-connected layers followed by a sigmoid and tanh activation functions, a concatenation operation, and another fully-connected layer followed by a sigmoid function. The output is a 1-D attention map of size 973 with values between 0 and 1, indicating the importance of the corresponding cluster to the classification task. Next, we identify the most predictive anatomical WM tracts. We first compute the mean importance of each cluster across all testing subjects. Then we find the top $T$ ($T$ = 50 in our experiment) clusters with the highest mean importance values. Finally, we identify all tracts to which the top 50 clusters belong, according to the ORG atlas [17].

## 2.4 Implementation details

All experiments are performed on a NVIDIA RTX A4000 GPU using Pytorch (v1.12.1) [28]. For the overall architecture, we use two EdgeConv layers and one 1-D convolutional layer to extract features. The two EdgeConv layers compute edge features with two fully-connected layers (64 , 64). Shortcut connections are included to extract multi-scale features and one 1-D convolutional layer (kernel size=1, output channel=64) to aggregate multi-scale features, where we concatenate features from previous layers to get a 64+64=128 dimension feature. After that, a flatten operation and two fully-connected layers follow to obtain the final classification results. Our overall network is trained for 200 epochs with a learning rate of 1e-5. The batchsize of training is 32 and Admax [29] is used for optimization. Source code will be made available.

## 3. RESULTS AND DISCUSSION

Four metrics are adopted in our study to evaluate sex classification performance: accuracy, precision, recall and F1 score. For precision, recall and F1 score, the averaged values of the two classes are calculated for evaluation.

### 3.1 Sex prediction performance

We compared the performance of our proposed method with two methods: SVM and 1D CNN [15]. In addition, an ablation study was performed to investigate the performance of our model without the attention module (TractGraph CNN w/o Att. Mod.). The sex classification results of all testing subjects from the ABCD and HCP datasets are shown in Tables 1 and 2.

Table 1. Comparison of sex classification performance across different methods in the ABCD dataset.

| Methods | SVM | 1-D CNN | TractGraph CNN w/o Att. Mod. | | TractGraph CNN | |
|---|---|---|---|---|---|---|
| | | | WMG | GMG | WMG | GMG |
| *Acc* | 73.46 | 82.77 | 85.13 | 84.59 | **85.50** | 84.80 |
| *Precision* | 73.21 | 82.83 | 85.09 | 84.58 | **85.46** | 84.77 |
| *Recall* | 73.23 | 82.88 | 85.11 | 84.52 | **85.49** | 84.79 |
| *F1* | 73.22 | 82.77 | 85.10 | 84.55 | **85.48** | 84.78 |

Table 2. Comparison of sex classification performance across different methods in the HCP dataset.

| Methods | SVM | 1-D CNN | TractGraph CNN w/o Att. Mod. | | TractGraph CNN | |
|---|---|---|---|---|---|---|
| | | | WMG | GMG | WMG | GMG |
| *Acc* | 90.673 | 93.229 | 93.750 | 94.271 | **94.792** | 93.229 |
| *Precision* | 90.506 | 93.229 | 93.760 | 94.326 | **94.791** | 93.245 |
| *Recall* | 91.414 | 93.234 | 93.760 | 94.254 | **94.791** | 93.259 |
| *F1* | 90.601 | 93.228 | 93.750 | 94.269 | **94.791** | 93.229 |

Generally speaking, our TractGraphCNN model with WMG shows the best performance across all compared methods. This indicates the strong potential of anatomically informed graphs to improve performance in deep learning tasks related to the brain's WM connections. Furthermore, we note that in general the WMG outperformed the GMG. This is likely because the neighborhoods constructed using fiber distances were able to capture more localized information. In comparison, many larger neighborhoods were induced when considering FS parcels. This could be seen in the neighborhood size, where in WMG each node (cluster) had 20 edges, while the number of edges per node in GMG ranged from 3 to 180. Overall, both TractGraphCNN approaches had good performance, in comparison with typical sex prediction accuracies across different MRI modalities and datasets ranging from 80-90% [10]. Note that a comparable recent study of sex prediction from HCP structural connectivity data achieved 92.75% accuracy [30]. Finally, despite the larger size of the ABCD dataset, all methods had much higher accuracy in the HCP dataset, likely related to the different neurodevelopmental stages of the subjects in the two datasets [31].

### 3.2 Interpretation of important tracts

Fig. 3 shows important tracts for the sex prediction task that were consistently identified across both ABCD and HCP experiments, for each graph type. We can observe that widespread regions in the WM are predictive of the sex of an individual. Three tracts (the bilateral cingulum and the left middle longitudinal fasciculus) were consistently predictive of sex across both graph types (WMG and GMG) and across both large datasets. Other interpretation results varied across graph types, indicating that the different graph structures helped the network focus on different informative brain connections. This further suggests the potentially complementary nature of the two investigated graphs, and the potential for future investigations into simultaneously leveraging multiple sources of anatomical information in network construction.

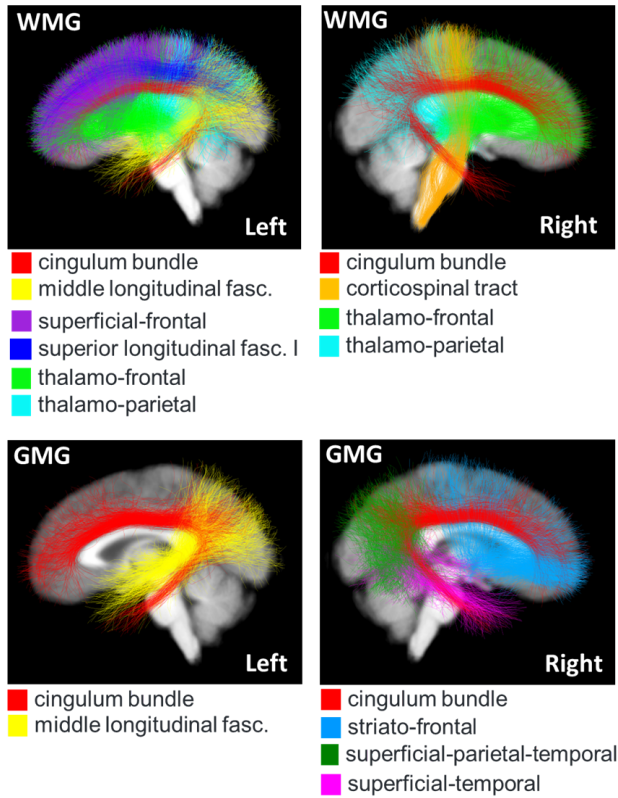

Fig. 3. Interpretation results of important tracts common across both ABCD and HCP datasets.

### 4. CONCLUSION

In this study, we proposed a novel anatomically informed graph CNN framework, TractGraphCNN, for machine learning using diffusion MRI tractography. The framework incorporates EdgeConv modules to aggregate features from white matter connections that are anatomically related, and an attention module that enables the interpretation of white matter tracts that are important for prediction. The results in a sex prediction testbed task demonstrated strong performance of TractGraphCNN in two large datasets. We found that white-matter-centric graphs were most successful overall. Across both datasets and both graph types, the bilateral cingulum and left middle longitudinal fasciculus were most predictive of the sex of an individual. Overall, this work shows the potential of incorporating sources of anatomical information, especially known anatomical similarities between input features, to guide convolutions in neural networks.

## 5. COMPLIANCE WITH ETHICAL STANDARDS

This study uses public HCP imaging data and no ethical approval was required. Approval was granted by the BWH IRB for use of the public ABCD data.

## 6. ACKNOWLEDGMENTS

We acknowledge the following NIH grants: P41EB015902, R01MH125860 and R01MH119222.